\documentstyle[aps,epsfig,twocolumn,citesort,psfig19,floats]{revtex}
\renewcommand {\baselinestretch}{1.0}

\newcommand{\ageq}{\mbox{\
\raisebox{-.9ex}{$\stackrel{\textstyle >}{\sim}$}\ }}

\def\x{{\mbox{\boldmath$x$}}}
\def\u{{\mbox{\boldmath$u$}}}
\def\r{{\mbox{\boldmath$r$}}}

\def\v{{\mbox{\boldmath$v$}}}

\def\rel{{Re_\lambda}}

\def\begineq{\begin{equation}}
\def\endeq{\end{equation}}
\def\be{\begin{equation}}
\def\ee{\end{equation}}

\begin{document}
\bibliographystyle{prsty}

\title{
Scaling of the irreducible SO(3)-invariants of velocity
correlations in
turbulence
}
\author{Siegfried Grossmann,
Detlef Lohse, and
Achim Reeh}
\address{
Fachbereich Physik der Universit\"at Marburg,
Renthof 6, D-35032 Marburg}

\date{\today}

\maketitle
\begin{abstract}
The scaling behavior of the SO(3) irreducible amplitudes $d_n^l(r)$ of
velocity structure tensors
(see L'vov, Podivilov, and Procaccia, Phys.\ Rev.\ Lett.\ (1997))
is  numerically examined
for Navier-Stokes turbulence. Here, $l$ characterizes the irreducible
representation by the index of the corresponding Legendre polynomial, and $n$
denotes the tensorial rank, i.e., the order of the moment. For moments
of different order $n$ but with the {\it same} representation index $l$
extended self similarity (ESS) towards large scales is found.
Intermittency seems to increase with $l$. 
We estimate
that a crossover behavior between {\it different}
 inertial subrange scaling
regimes 
in the longitudinal and transversal structure functions 
will hardly be detectable 
for achievable Reynolds numbers. 
\end{abstract}

\vspace{1cm}



The most fundamental objects to analyze the structure of turbulent velocity
fields $\u(\x,t)$ are the tensorial moments of the velocity differences
$v_i(\r;\x,t) = u_i(\x+\r,t) - u_i(\x,t)$,
averaged over time $t$ or/and position $\x$, considered as functions of
scale $\r$, 
\be
D_{i_1,i_2,\ldots,i_n}(\r) = \langle v_{i_1} v_{i_2} \cdots v_{i_n} \rangle
\label{eq1}
\ee
If the eddy size $r = |\r|$ is in the inertial subrange (ISR), i.e.,
$\eta \ll r \ll L$, algebraical scaling of the moments is expected.
Here, $\eta$ is the inner (Kolmogorov) scale and $L$ the external length scale
\cite{my75}. 
If the turbulent flow field can be 
considered as statistically isotropic (or close to),
one better uses rotational
invariants instead of the tensorial components, in order to cope with the
multitude of scaling exponents. The most commonly used invariants are the
structure functions of the 
longitudinal velocity component $v_L = \v \cdot \r^0$ and the 
transversal velocity $\v_T = \v - v_L \r^0$; here,
$\r^0$ denotes the unit vector in $\r$ direction. 
We denote these structure functions as 
\be
  D_n^L (r) =
  \langle [ v_L(\r;\x,t)  ]^n \rangle
\propto r^{\zeta_n^L}, 
\label{eq2}
\ee
\begin{equation}
  D_n^T (r) =
  \langle |  \v_T(\r;\x,t) |^n \rangle 
\propto r^{\zeta_n^T};
\label{eq3}
\end{equation}
both are assumed to scale in the ISR with the corresponding
exponents $\zeta_n^L$ and $\zeta_n^T$. 
A third convenient structure function is the $n$-th order moment of the
{\it modulus}
of the eddy velocity difference $\v(\r;\x,t)$ which again is assumed
to scale 
\begin{equation}
  D_n^M (r) =
  \langle  |\v(\r;\x,t)|^{n} \rangle \propto r^{\zeta_n^M}.
\label{eq4}
\end{equation}
Traditionally, it was believed that all
three scaling exponents are the same, $\zeta_n = \zeta_n^M = \zeta_n^L
=\zeta_n^T$. 
But recent advances in experimental technology
\cite{sad94,her95,nou97,dhr97} and computational power
and technique \cite{bor97,gro97a} raised increasing
 doubts if this is true
for general moments of
 order $n$, as it is for the most often considered 2nd order
structure function, $n=2$, where the condition of incompressibility
enforces $D_2^L \propto D_2^T \propto D_2^M \propto r^{\zeta_2}$.
For general $n$,
 it was found in several experiments and simulations
that the
degree of intermittency (i.e., the deviations of the scaling exponents from the
classical value $\zeta_n = n/3$) is considerably larger in the transversal
moments compared to the longitudinal ones; for a summary of the results see
table 1 of ref.\ \cite{gro97a}. 

In a recent paper, L'vov, Podivilov, and Procaccia \cite{lvo97} suggested that
it were not the longitudinal and the transversal structure functions which
obey clean algebraic scaling, but the amplitudes of the moment tensor
eq.\ (\ref{eq1}) 
decomposed into the irreducible representations of the rotation group SO(3),
\be
d_n^l (r) \propto 
  \langle (\v^2(\r;\x,t))^{n/2}  P_l ( \v^0 \cdot \r^0 ) \rangle 
\propto r^{\zeta_n^l} .
\label{eq5}
\ee
The representation label $l$ runs through $0 \le l \le n$ with the same
parity as $n$, if statistical reflection symmetry of the turbulent flow field
is guaranteed \cite{lvo97}; 
$P_l$ is the Legendre polynomial. 
The amplitude 
of the unity representation, $d^0_n(r)$,
is already part of the conventional set of structure function, since
$d^0_n(r) \propto D_n^M(r)$.

For the second and fourth order structure tensors the amplitudes  $d_2^l(r)$
and $d_4^l(r)$ are linear combinations of the longitudinal,  transversal, and
modulus structure functions. We follow L'vov et al.'s  definitions \cite{lvo97}
$a_0 = d_2^0$,
$a_2 = d_2^2$,
$c_0 = d_4^0$,
$c_2 = d_4^2$,
$c_4 = d_4^4$
obtaining
\be
 \left( \begin{array}{c}
a_0 \\ a_2
\end{array}
\right)
=
 \left( \begin{array}{cc}
{1\over 3} & 0  \\ {1\over 6}  & -{1\over 2}  
\end{array}
\right)
 \left( \begin{array}{c}
D_2^M \\ D_2^L
\end{array}
\right),
\label{eq6}
\ee
\be
 \left( \begin{array}{c}
c_0 \\ c_2 \\ c_4
\end{array}
\right)
=
 \left( \begin{array}{ccc}
{1 \over \sqrt{5}} & 0 & 0 \\
{1\over 2 \sqrt{7}}  & {3\over 2\sqrt{7}} &  -{3\over 2\sqrt{7}}
 \\
 {-6\over \sqrt{70}} & {10 \over \sqrt{70}} & {15 \over \sqrt{280}}  
\end{array}
\right)
 \left( \begin{array}{c}
D_4^M \\ D_4^L \\ D_4^T
\end{array}
\right).
\label{eq7}
\ee
On the rhs also other ways of representing the n-rank
velocity correlation tensor can alternately be given, using e.g.\ $D_2^T$ in
(\ref{eq6}) or mixed transversal/longitudinal moments in (\ref{eq7})
as done in eq.\ (13.81) of ref.\ \cite{my75} or in ref.\ \cite{oul96}
which uses
$D_{11} = D_2^L$, $D_{22} = D_2^T/2$, 
$D_{1111} = \langle v_1^4 \rangle =D_4^L$,
$D_{1122} = \langle v_1^2 v_2^2\rangle$, and 
$D_{2222} = \langle v_2^4 \rangle =3D_{2233} = 3D_4^T/8$, where the
1-axis has been put in the longitudinal direction parallel to $\r$. For
these structure functions we obtain
\be
 \left( \begin{array}{c}
c_0 \\ c_2 \\ c_4
\end{array}
\right)
=
 \left( \begin{array}{ccc}
{1 \over \sqrt{5}} & {4\over \sqrt{5}} & {8\over 3\sqrt{5}} \\
{2\over  \sqrt{7}}  & {2\over \sqrt{7}} &  -{8\over 3 \sqrt{7}}
 \\
 \sqrt{8\over 35} & -12\sqrt{2 \over 35} & \sqrt{8 \over 35}  
\end{array}
\right)
 \left( \begin{array}{c}
D_{1111} \\ D_{1122} \\ D_{2222}
\end{array}
\right).
\label{c_mat}
\ee

The point of L'vov et 
al.\ is that the invariants $a_l$, $c_l$ on the lhs of
eqs.\ (\ref{eq6}), (\ref{eq7}), and (\ref{c_mat})
 are {\it distinguished} because
the $d_n^l(r)$ are the amplitudes of the structure tensor for its
decomposition 
into the components of the {\it irreducible representations} of the
rotational symmetry group SO(3).

In this paper we present the scaling properties of the
{\it fourth} order moments $d_4^l(r)$
from a full numerical simulation of the Navier-Stokes equation on a
$96^3$ grid
with periodic boundary conditions. The numerical turbulence is forced on the
largest scales, the averaging time is about 120 large eddy turnovers, and
the Taylor-Reynolds number is $\rel = 110$.
The isotropy of the flow has carefully been checked; for details of the
simulation we refer to ref.\ \cite{gro97a}.

\begin{figure}[htb]
\setlength{\unitlength}{1.0cm}
\begin{picture}(6,6.6)
\put(-0.5,0.5)
{\psfig{figure=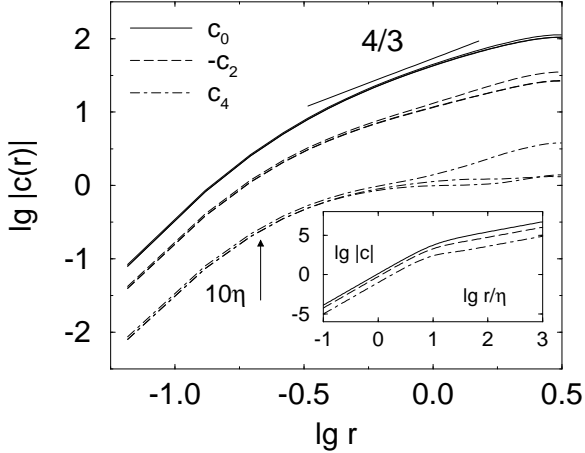,width=8cm,angle=-90}}
\end{picture}
\caption[]{
Fourth order structure functions
$c_{0,2,4}(r)$ as functions of $r$.
As tested for smaller Reynolds numbers and longer averaging times,
the wiggle in $c_4(r)$ at large $r$ is not statistically safe.
It seems that very long averaging times are necessary for 
moments with large $l$ to converge at large scales.
The inset shows $c_{0,2,4}(r)$ (top to bottom)
as they follow from the
Batchelor parametrization (\ref{bat_beide})
and eqs.\ (2) of ref.\ \cite{oul96}.
The different magnitudes and the distinct transitional behavior 
of the different irreducible representations can be recognized. 
Note that the local slope of $\lg c_4(r)$ vs $\lg r$
around the transition is
{\it not} monotonous. -- 
By assumption the ISR scaling exponents are the same. 
}
\label{fig1}
\end{figure}

The {\it second} order moments all asymptotically scale the same because
of incompressibility.
Assuming classical scaling $\zeta_2 = 2/3$
one obtains $D_2^L = 4D_2^T/3$
and $a_0 = 11 a_2 = D_2^M/3$.
In figure \ref{fig1}
we give the 
{\it fourth} order structure functions
$c_{0,2,4}(r)$. As expected for this low $\rel$, the scaling properties of
these structure functions is very poor, because there is not yet a well
developed ISR. There is analytical behavior $\propto r^4$ in the viscous
subrange (VSR)
followed by a transition and leveling off in the inertial and stirring
subrange around $r \sim L$. 
What can be said, however, is that
with increasing $l$
(i) the magnitude of $c_l(r)$ decreases and (ii) 
the degree of intermittency seems to increase,
$
\zeta_4^4 < 
\zeta_4^2 < 
\zeta_4^0 < 
4/3$.
The reason for (i) is that $c_0$ is a {\it sum} of positive definite
structure functions, whereas $c_2$ and the more
$c_4$  are {\it differences} thereof, similar to $a_2 = (D_{22}-D_{11})/3$
which also is much smaller than $a_0=D_{11} +2 D_{22}$. The reason for (ii)
presumably is that larger $l$ in (\ref{eq5}) means
the probing of smaller scale structures which are traditionally associated with
stronger intermittency.

\begin{figure}[htb]
\setlength{\unitlength}{1.0cm}
\begin{picture}(6,11)
\put(0.0,6.5)
{\psfig{figure=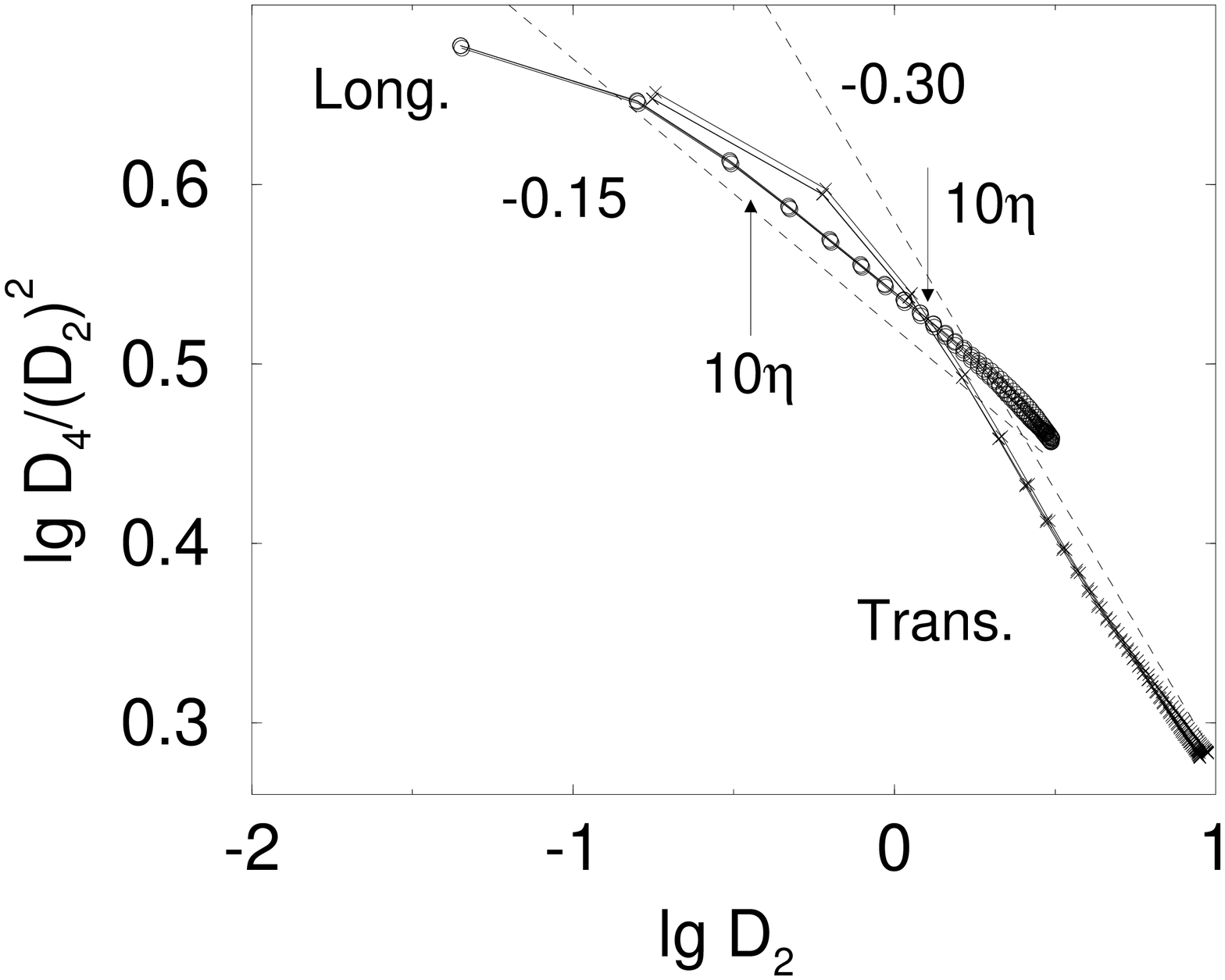,width=7cm,angle=-90}}
\put(0.5,0.5)
{\psfig{figure=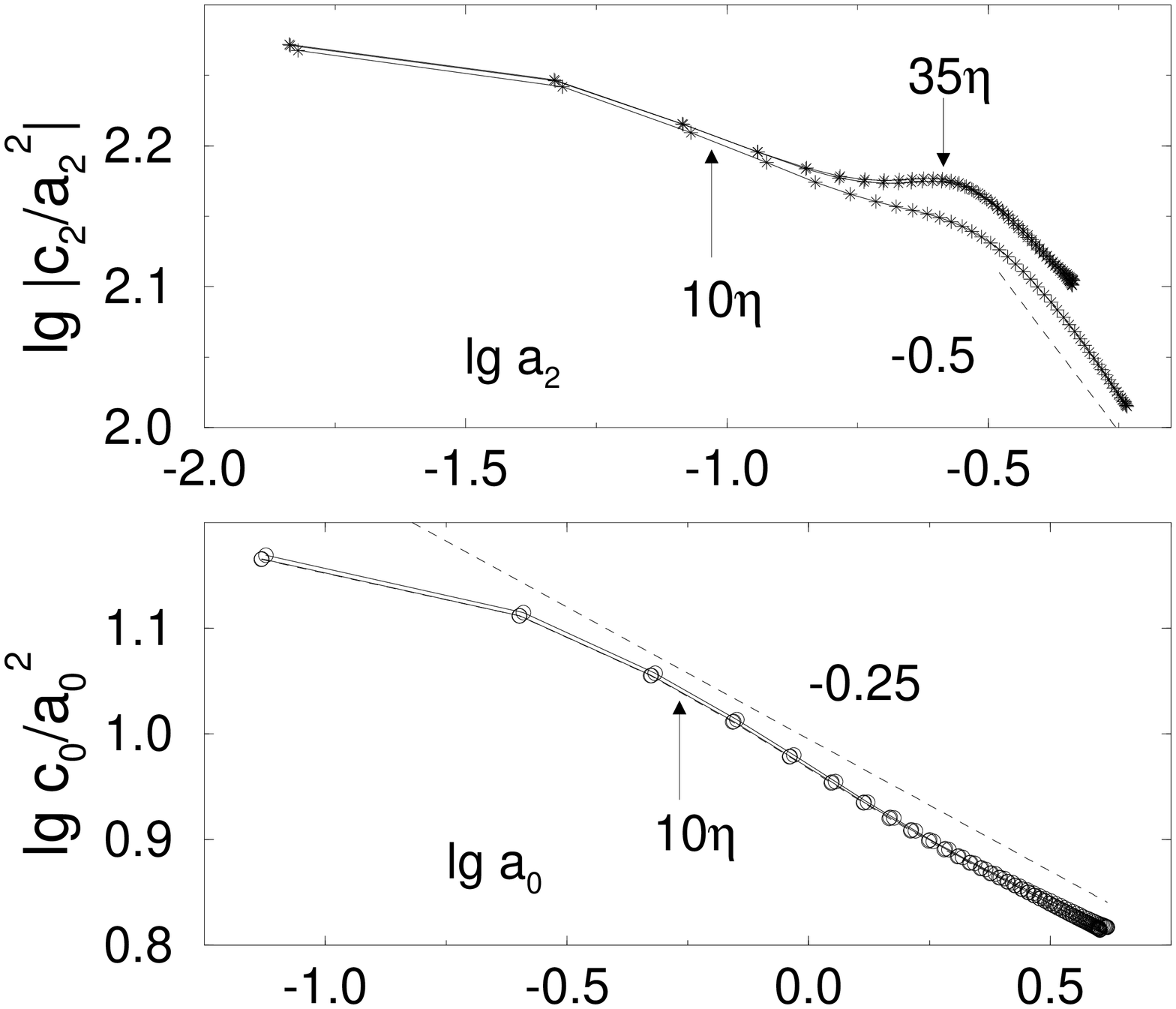,width=7cm,angle=-90}}
\end{picture}
\caption[]{
Compensated ESS plots for different space directions
of fourth vs second order structure functions,
giving the ISR scaling exponent
$\rho_i = (\zeta_4^i - 2\zeta_2^i) / \zeta_2^i$.
The plot $c_2/a_2^2$ vs $a_2$ displays a distinct bump around $35\eta$
for all three space directions, though it is slightly differently
developed in strength, possibly because of the unavoidable anisotropy in
the forcing, possibly because of the too short averaging time
(Fig.\ 2b, upper).
 -- Convergence is much less a problem
in the ESS plot $c_0/a_0^2$ vs $a_0$ (Fig.\ 2b, lower)
and consequently also in the
ESS plots of the longitudinal and transversal structure functions where
the rank zero contributions dominate (Fig.\ 2a).
}
\label{fig2}
\end{figure}

Fortunately, the extended self similarity method
(ESS, \cite{ben93b})
allows for more quantitative statements. Here, we focus on the scaling
of the fourth order structure functions vs second order ones. More
specifically, to visualize the deviations from classical scaling we calculate
{\it compensated} ESS plots
$D_4^i/ (D_2^i)^2$ vs $D_2^i$, $i=L,T,M$, and 
$d_4^l/ (d_2^l)^2$ vs $d_2^l$, $l=0,2$,
see figure \ref{fig2}.
For $l=0$ we find ESS scaling from $r\sim 10\eta$ up to $r\sim L$, resembling
the ESS scaling for the longitudinal and transversal structure functions
figure \ref{fig2}a which was extensively analyzed in 
\cite{ben93b,gro97b}. 
For $l=2$ we find ESS towards large scales  $r\ageq 50\eta$, but
{\it no ESS towards smaller scales } $r< 50\eta$. Instead, there is a bump in
the curve $c_2 / a_2^2 $ vs $a_2$ for $r\sim 36\eta$. We checked very carefully
whether the bump would smooth for increasing averaging time. This is
{\it not} the
case. It also persists for different type of large scale forcing  and
smaller 
Reynolds number, but much larger averaging time.

At first sight, the bump was a surprise to us. However,
we suggest that it can be understood as
a transition phenomenon from the 
VSR to the ISR, similar to the one seen in
figure 3 of ref.\ \cite{gro97b}. Hitherto, it was not observed in ESS plots of
longitudinal and transversal structure functions as both are dominanted by the
rank
zero contribution $d_n^0$
 which does show ESS.

\begin{figure}[htb]
\setlength{\unitlength}{1.0cm}
\begin{picture}(6,9)
\put(-0.8,1.0)
{\psfig{figure=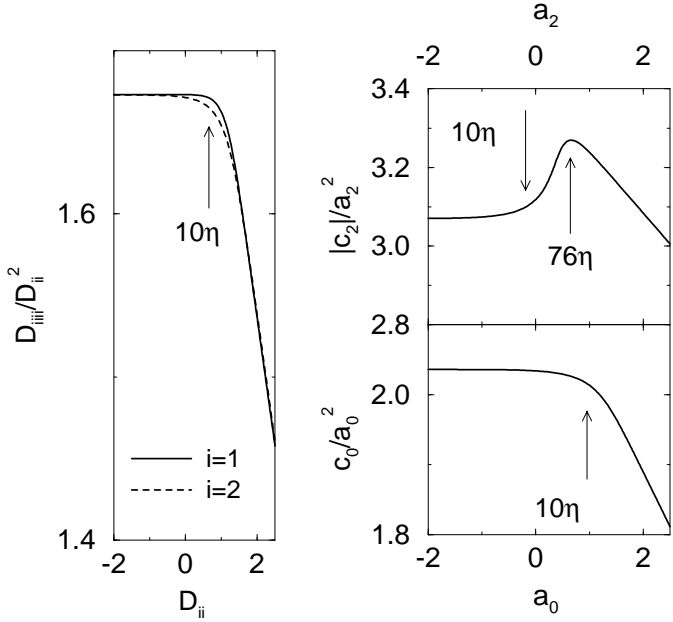,width=10cm,angle=-90}}
\end{picture}
\caption[]{
Compensated ESS plots
following from the Batchelor parametrizations (\ref{bat_beide})
and eqs.\ (2) of ref.\ \cite{oul96}. 
The curve 
$c_2/a_2^2$ vs $a_2$ is the only ESS plot which is not dominanted by
the tensors of rank zero and it does show a bump at the VSR-ISR
transition. That the ISR scaling exponent $\rho_i$ is the same for
the four curves shown here for demonstrative reasons
is a trivial consequence of the closure assumptions
of ref.\ \cite{oul96}, but does not hold for the real data. 
}
\label{ess_bat}
\end{figure}

For further support of this interpretation, we parametrize the
$d_n^0(r)$ 
within Batchelor's parametrization \cite{my75},
\be
d_n^0(r) 
\propto D_n^M (r) \propto   {r^n 
\left[ 1+ \left( {r\over r_c}\right)^2\right]^{(-n+\zeta_n)/2} },
\label{bat_beide}
\ee
$n=2,4$, with the She-Leveque model \cite{she94} values 
$\zeta_2 = 0.70$, 
$\zeta_4 = 1.28$, and $r_c = 10\eta$. Incompressibility gives
$D_2^L (r) =  r^{-3}  \int_0^r D_2^M(\tilde r) \tilde r^2 d\tilde r$
and via eq.\ (\ref{eq6}) all other second order structure functions follow.
For the fourth order moments, an analogous  relation
does not exist. However, within some closure approximations \cite{oul96}
(whose nature is  controversal),
all 4th order structure functions follow from $D_4^L(r)$, cf.\
eq.\ (2) of ref.\ \cite{oul96}.
We stress that those relations are not generally
true and their consequence
that all 4th order structure functions  scale the   
same is in direct contradiction to our and others' findings.
However, for the
demonstration of transitional effects, for which the
different intermittency in
the ISR does not matter, eqs.\ (2a) -- (2c)
of ref.\ \cite{oul96} could be  useful. Employing them we derive an
ODE for $D_{1111} = D_4^L$,
\be
D_4^M = 5D_4^L + {11\over 6} r {d \over dr} D_4^L
+ {1\over 6} r^2 {d^2 \over dr^2 } D_4^L
\label{incom4}
\ee
which can numerically be solved for the given $D_4^M(r) $ 
of eq.\ (\ref{bat_beide}).
With \cite{oul96}
\be
D_{1122} = {1\over 3} D_4^L + {r\over 12} {d\over dr} D_4^L,
\label{eq1122}
\ee
\be
D_{2222} = D_4^L +{9\over 16} r {d\over dr} D_4^L
+ {1\over 16} r^2 {d^2 \over dr^2} D_4^L
\label{eq2222}
\ee
and eqs.\ (\ref{eq7}) --
 (\ref{c_mat}) all other fourth order moments follow. The resulting
compensated ESS plots are shown in figure \ref{ess_bat}.
Indeed, in the ESS plot $c_2/a_2^2$ vs $a_2$ a bump occurs at the VSR-ISR
transition, similar to what we found in the numerical simulation. All other
shown compensated ESS plots are dominated by
$D_4^M = \sqrt{5}c_0$ and $D_2^M = 3a_0$, and therefore display
ESS, as $c_0$ vs $a_0$ does. After the
functional shape of the ESS transition from VSR to ISR
in $D_4^M$ vs $D_2^M$ is now believed to be rather universal, it would be worth
while to analyze in various experimental and numerical flows, whether the
first angular contribution, i.e., $c_2$ vs $a_2$, also is
somehow universal and thus 
displays the type of structure as we found in figure \ref{fig2}b.

We now come back to the numerical results and focus on the ESS 
scaling exponents  of figure \ref{fig2} which we denote by 
$\rho_i = (\zeta_4^i - 2\zeta_2^i) / \zeta_2^i$, 
$i=L,T,M$ or $i=0,2$. The deviation of the $\rho_i$ from zero characterizes the
degree of intermittency of the corresponding moment. We find
$\rho_L = -0.15$,
$\rho_T = -0.30$ and 
$\rho_M =\rho_0 = -0.25$,
$\rho_2 = -0.5$, again showing that the degree of intermittency
is higher in the
$d_n^l$ with larger $l$. The She-Leveque model value
(with the original She-Leveque parameters adopted to the longitudinal
structure function) 
\cite{she94}
for $\rho$ is $\rho= -0.16$.

We checked the possibility of scaling behavior if amplitudes corresponding to
different irreducible subspaces are mixed:
We do {\it not} find ESS if we plot structure functions
$d_4^l/ (d_2^{l^\prime})^2$
vs $d_2^{l^\prime}$ with {\it different } $l\ne l'$.

It will not have escaped the reader's attention that the
simultaneous assumption of pure
scaling behavior of both the $D_4^{L,T,M}$ as well as the $c_4^{0,2,4}$ is 
self contradictory if the exponents with different $l$ are different.
We follow L'vov et al.'s argument that the $d_n^l$ are the more fundamental
structure
functions and (employing eq.\ (\ref{eq7}) and incompressibility) write
the ratios
$D_4^{L,T} / (D_2^{L,T})^2 $ as a sum of ratios of the $d_n^l$,
\be
{D_4^L\over (D_2^L)^2 }
\propto {c_0 \over a_0^2 } + 2 \sqrt{5\over 7} {c_2 \over a_0^2}
+\sqrt{8\over 7} {c_4 \over a_0^2},
\label{eq9}
\ee
\be
{D_4^T\over (D_2^T)^2 }
\propto {c_0 \over a_0^2 } - \sqrt{5\over 7} {c_2 \over a_0^2}
+ {3 \over 4} \sqrt{2\over 7} {c_4 \over a_0^2}. 
\label{eq10}
\ee
{}From our numerics (see fig.\ \ref{fig1})
the first term is found to be 
the leading order term. It
represents the scaling of the modulus structure functions
eq.\ (\ref{eq4}). In the first (and larger)
correction term the approximation
$a_0 \approx 11a_2$ (resulting from  $\zeta_2 \approx 2/3$ and
incompressibility) can be made, leaving only ratios
whose scaling
we can determine from the ESS-plots figure \ref{fig2}b; 
(the $c_4$-term hardly contributes for large $r$).
With that approximation
the qualitative features of figure \ref{fig2}a can be understood from
eqs.\ (\ref{eq9}) and (\ref{eq10}): As $c_2(r) < 0$ the
$c_2/a_2^2$ correction term to the leading 
$c_0/a_0^2$ term is negative [positive] for
$D_4^L / (D_2^L )^2 $ 
[$D_4^T / (D_2^T )^2 $], leading to a less steep [steeper] ``apparent''
slope for the ESS exponents $\rho_L = -0.15$ [$\rho_T = -0.30$]
of the longitudinal [transversal] structure function compared to the
leading
contribution with $\rho_0 = \rho_M = -0.25$. Even that the correction to
$\rho_0 = -0.25$ is twice as big for the longitudinal structure function than
for the transversal one can be seen from eqs.\ (\ref{eq9}) and
(\ref{eq10}).

Finally, we would like to estimate for what Reynolds number
two distinct scaling regimes  (in $r$)
may be observable in $D_4^{L,T}$.
Therefore, we plug in the scaling laws
\be
{c_i\over a_i^2}(r) = {c_i\over a_i^2} (L) \left( {r\over L
}\right)^{\zeta_2\rho_i}; \qquad i=0,2
\label{sca}
\ee
and obtain with the numerical values at $r=L$, 
$c_0/a_0^2 \approx 6$ and 
$c_2/a_2^2 \approx -100$,
\be
r^{\zeta_2 \rho_L } \propto {D_4^L\over (D_2^L)^2 }
\propto r^{\zeta_2 \rho_0} \left( 1-\alpha \left({r\over
L}\right)^{\zeta_2(\rho_2-\rho_0)} +c_4\hbox{-corr.}\right) ,
\label{eq11}
\ee
\vspace{-1cm}
\be
r^{\zeta_2 \rho_T } \propto {D_4^T\over (D_2^T)^2 }
\propto r^{\zeta_2 \rho_0} \left( 1+{1\over 2} \alpha \left({r\over
L}\right)^{\zeta_2(\rho_2-\rho_0)} +c_4\hbox{-corr.}\right) .
\label{eq12}
\ee
We get $\alpha \sim 0.2$,
$\zeta_2 (\rho_2 - \rho_0 ) = {2\over 3} (-0.5 + 0.25) \approx -0.17$.
Note that for small enough $r$ the second term in (\ref{eq12}) may
dominate the first one and for even smaller $r$ the third term
will contribute. [In eq.\ (\ref{eq11}) the situation is more complicated as the
second term has negative sign, but the lhs is positive definite.]
Therefore, in principle $D_4^T/(D_2^T)^2$ shows several different scaling
regimes. However, it will be very hard to detect these different regimes as the
required span of the Reynolds number is too large. In eq.\ (\ref{eq12}) 
the ratio $L/r$ has to be as large as $L/r =
(2 /  \alpha )^{1/0.17} \sim 10^{6}$
for the second term to overtake the first one. 
We put $r=\eta$ and estimate that this means $Re\sim 10^{8}$,
which is hard to achieve in today's experimental or numerical flows.
What shall be detectable if L'vov et al.'s conjecture \cite{lvo97} is right 
is that the apparent scaling exponents of the structure functions
$D_n^{L,T}(r)$ or ESS scaling exponents thereof are slightly $Re$ dependent 
whereas
the scaling exponents of the irreducible objects $d_n^L(r)$
or their ESS exponents (the exponents of plots $|d_n^l(r)|$ vs
$|d_m^l(r)|$) might well be universal, i.e., Reynolds number independent.
 Measuring such exponents
up to very large Reynolds numbers \cite{dhr97} seems to be of prime
importance for our further understanding of fully developed
turbulence.

\vspace{0.3cm}

\noindent
{\bf Acknowledgments:}
We thank
I.\ Procaccia and K.\ R.\ Sreenivasan for very helpful exchange.
Support for this work by
the DFG under grant SBF185-D3 and by
the German-Israel Foundation (GIF) is acknowledged, and also by the
HLRZ J\"ulich supplying us with the necessary computer time.


\vspace{-0.95cm}


\end{document}